\documentclass[twocolumn,showpacs,showkeys,superscriptaddress,longbibliography,floatfix]{revtex4-1}
\usepackage{graphicx}
\usepackage{amsmath}
\usepackage{multirow}

\date{\today}

\begin{document}

\author{Unnar B. Arnalds}
\affiliation{Science Institute, University of Iceland, Dunhaga 3, IS-107 Reykjavik, Iceland}
\email{uarnalds@hi.is}

\author{Jonathan Chico}
\affiliation{Department of Physics and Astronomy, Uppsala University, SE-75120 Uppsala, Sweden}

\author{Henry Stopfel}
\affiliation{Department of Physics and Astronomy, Uppsala University, SE-75120 Uppsala, Sweden}

\author{Vassilios Kapaklis}
\affiliation{Department of Physics and Astronomy, Uppsala University, SE-75120 Uppsala, Sweden}

\author{Oliver B\"arenbold}
\affiliation{Department of Physics and Astronomy, Uppsala University, SE-75120 Uppsala, Sweden}

\author{Marc A. Verschuuren}
\affiliation{Philips Research Laboratories, Eindhoven, Netherlands}

\author{Ulrike Wolff}
\affiliation{IFW Dresden, Institute of Metallic Materials, D-01171 Dresden, Germany}

\author{Volker Neu}
\affiliation{IFW Dresden, Institute of Metallic Materials, D-01171 Dresden, Germany}

\author{Anders Bergman}
\affiliation{Department of Physics and Astronomy, Uppsala University, SE-75120 Uppsala, Sweden}

\author{Bj\"{o}rgvin Hj\"{o}rvarsson}
\affiliation{Department of Physics and Astronomy, Uppsala University, SE-75120 Uppsala, Sweden}

\title{A new look on the two-dimensional Ising model: thermal artificial spins}

%\keywords{Magnetic ordering, artificial spins, Ising model}

\begin{abstract}
We present a direct experimental investigation of the thermal ordering in an artificial analogue of an asymmetric two dimensional Ising system composed of a rectangular array of nano-fabricated magnetostatically interacting islands. During fabrication and below a critical thickness of the magnetic material the islands are thermally fluctuating and thus the system is able to explore its phase space. Above the critical thickness the islands freeze-in resulting in an arrested thermalized state for the array. Determining the magnetic state of the array we demonstrate a genuine artificial two-dimensional Ising system which can be analyzed in the context of nearest neighbour interactions.
\end{abstract}

\maketitle

\begin{figure}[t]
\begin{center}
  \includegraphics[scale=1]{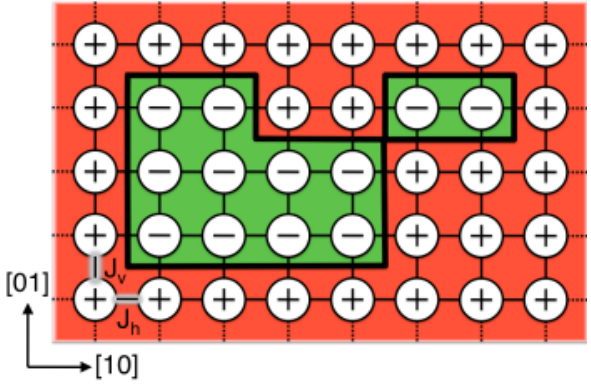} \\
\end{center}
\caption{The two dimensional Ising system. A two dimensional Ising system consists of an array of spins which are only capable of pointing in two opposite directions, ${\bf s}=\pm 1$ and interacting with their four nearest neighbours with interaction energies $J_h$ and $J_v$ along the [10] and [01] lattice directions. At $T=0$ K the ground state is doubly degenerate consisting of all spins pointing in the (+) or the (-) directions for the case of a symmetric Ising system. For $T>0$ K defects above the ground state occur enclosed by a domain wall separating the areas of spins in the all (+) or all (-) directions. 
}
\label{fig:2d_Ising}
\end{figure}

%%%%%%%%%%%%%%%%%%%%%%%%%%%%%%%%%%
%\noindent {\bf Introduction} \\
%%%%%%%%%%%%%%%%%%%%%%%%%%%%%%%%%%
%%%%%%%%%%%%%%%%%%%%%%%%%%%%%%%%%%
%\vspace{2ex}\noindent {\bf \large The two dimensional Ising model}\\
%%%%%%%%%%%%%%%%%%%%%%%%%%%%%%%%%%
\noindent The Ising model invented by Wilhelm Lenz and solved in one dimension by Ernst Ising in 1924 is one of the pillars of statistical mechanics\cite{ising, History_Ising_model_RevModPhys.39.883}.  Although built on a simple basis, that of an interacting system composed of a chain of entities with only two discrete states, ${\bf s}=\{1,-1\}$, the Ising model still to this day is used to model magnetic systems and can be applied to a wealth of atomistic and mesoscopic experimental systems ranging from ferromagnetic ordering and atomic-scale antiferromagnets \cite{Loth:Science2012} to the ordering of binary colloidal structures\cite{Khalil_NatComm_2012} and thermal artificial spin systems \cite{Zhang_Nat_2013}.
In the two dimensional case spins are arranged on a square lattice and each spin interacts with four nearest neighbours, as seen in Fig. \ref{fig:2d_Ising}. 
The total energy of such a two-dimensional Ising system can be described by the equation
\begin{equation}
E =  -  J_h \sum_{i,j}^h  {\bf s}_i\cdot {\bf s}_j -  J_v \sum_{i,j}^v  {\bf s}_i\cdot {\bf s}_j 
\label{eq:hamiltonian}
\end{equation}
where $J_h$ and $J_v$ correspond to interaction energies in the horizontal [10] and vertical [01] lattice directions of the two dimensional crystal and the sum is taken over all pairs of nearest neighbour spins ${\bf s}_i$ and ${\bf s}_j$. 
As opposed to the one dimensional model which shows no spontaneous magnetization at temperatures $T>0$ 
a spontaneous magnetization appears in the two dimensional case\cite{peierls} with an order parameter given by\cite{Onsager_2d_Ising,Yang:1952}
\begin{equation}
M(T) = \biggl( 1 - \biggl[\sinh \biggl(\frac{2J_h}{k_{\mathrm{B}}T}\biggr) \sinh \biggl(\frac{2J_v}{k_{\mathrm{B}}T} \biggr) \biggr]^{-2}   \biggr)^{\frac{1}{8}}
\label{eq:MofT}
\end{equation}
where $T$ corresponds to the temperature and $k_{\mathrm{B}}$ is the Boltzmann constant. The model is not complicated by the choice of different values or signs of $J_h$ and $J_v$ \cite{Onsager_2d_Ising}.
For $J>0$ the interaction between neighbouring spins favors a parallel alignment (see Fig. \ref{fig:2d_Ising}) while in the case of $J<0$ spins have a preference for an antiferromagnetic alignment.

Nano-patterned single-domain magnetic thin film islands have been a prominent candidate in recent years for creating artificial analogues of interacting systems. Using modern lithographic techniques it has become feasible to design directly the shape of such islands and their geometrical arrangement creating two dimensional arrays of interacting artificial spins. Artificial Ising-like spins can be realized by designing elongated islands of thin films materials, for which shape anisotropy confines the magnetization to only two possible orientations. By arranging such islands in different geometries a wealth of interacting systems can be studied including cellular automata \cite{Imre_Science_2006,Cowburn_Science_2000} and  frustrated systems such as artificial spin ice \cite{wang_nature,Nisoli_Review}. The two dimensional nano-scale nature of these systems enables their state to be directly determined by imaging techniques such as magnetic force microscopy\cite{wang_nature,Morgan_NPHYS}, photoemission electron microscopy\cite{mengotti:NPHYS2011,KagomeAPL,Kapaklis:NatNano:2014} and Lorentz microscopy \cite{Daunheimer:PRL:2011}. 

\begin{figure}[t]
\begin{center}
  \includegraphics[scale=1]{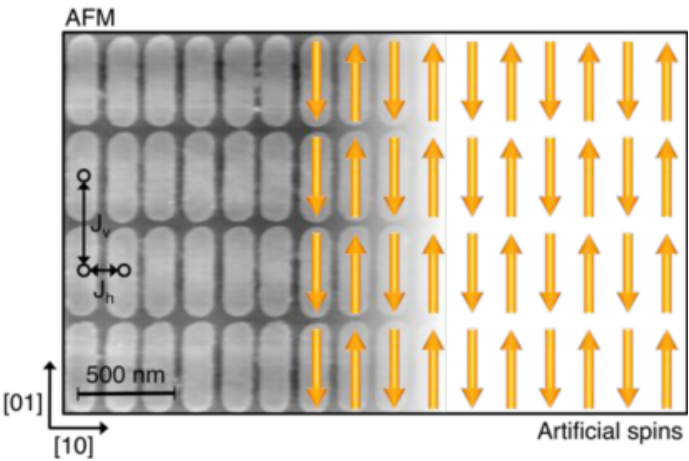}
\end{center}
\caption{Nano-patterned artificial spins. Atomic force microscopy (AFM) image of part of the patterned array showing the structure and arrangement of the nano-patterned islands (see Methods). The shape of the islands confines the magnetization to point along the long axis of the islands and during thermalization at the onset of a frozen state they can be considered as thermally active superspins. In the superspin model the vertical coupling ([01] direction) prefers a ferromagnetic alignment of the spins ($J_v>0$) whereas the lateral coupling ([10] direction) prefers an  antiferromagnetic alignment ($J_h<0$) resulting in the ground state ordering of the artificial spins illustrated to the right in the figure. 
}
\label{fig:AFM}
\end{figure}

In this paper we investigate a two dimensional array composed of elongated thin film islands in a square lattice (see Fig. \ref{fig:AFM}). 
During fabrication the array goes through a dynamic phase enabling the system to thermally explore its phase space leading to a low energy ordered state \cite{Morgan_NPHYS}. After this dynamic phase the state of the system becomes frozen-in, generating a snapshot of an arrested thermalized state. The determination of the magnetization of individual islands in the array allows us to demonstrate that the dynamic phase leads to an ordering of the magnetic state of the array, which can be described by the two-dimensional Ising model.

\begin{figure*}[t]
\begin{center}
  \includegraphics[scale=1]{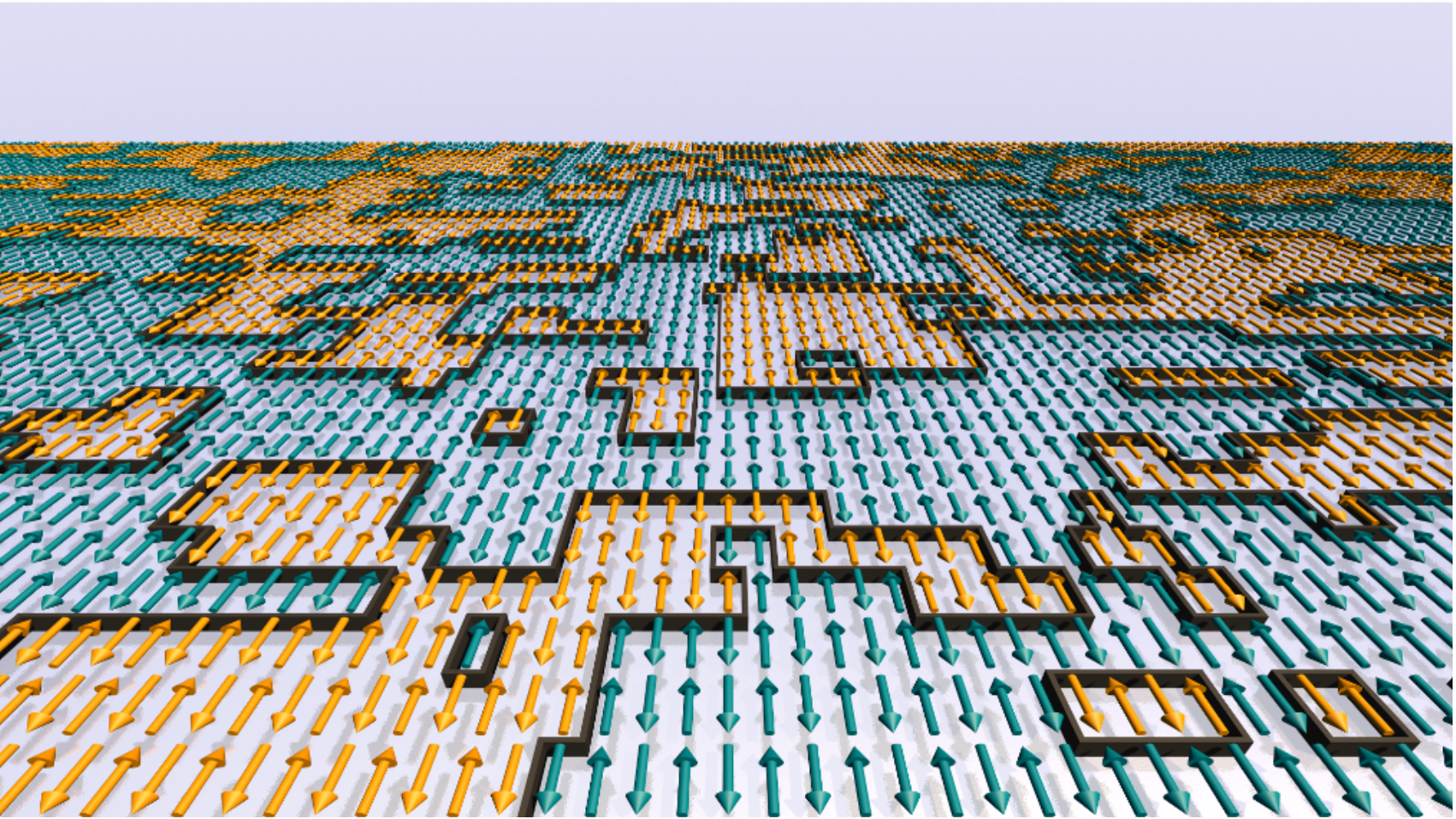}
\end{center}
\caption{Snapshot of an arrested state. The ordering of the investigated array illustrated through highlighting the two degenerate ground state and the domain walls separating them. As compared to a fully ferromagnetic 2-dimensional Ising system the antiferromagnetic coupling leads to an additional asymmetry in the population of the up and down islands themselves, which is indicative of the influence of an external field. Out of the $N=10487$ islands analyzed $n_u=5294$ were found to point in the up direction while $n_d=5193$ were found to point in the down direction resulting in an asymmetry within statistical limits. An overview of the entire investigated array as well as a data file containing the directions of all islands is supplied in Supplementary material. 
}
\label{fig:3dspins}
\end{figure*}

\begin{figure}[t]
\begin{center}
  \includegraphics[scale=1]{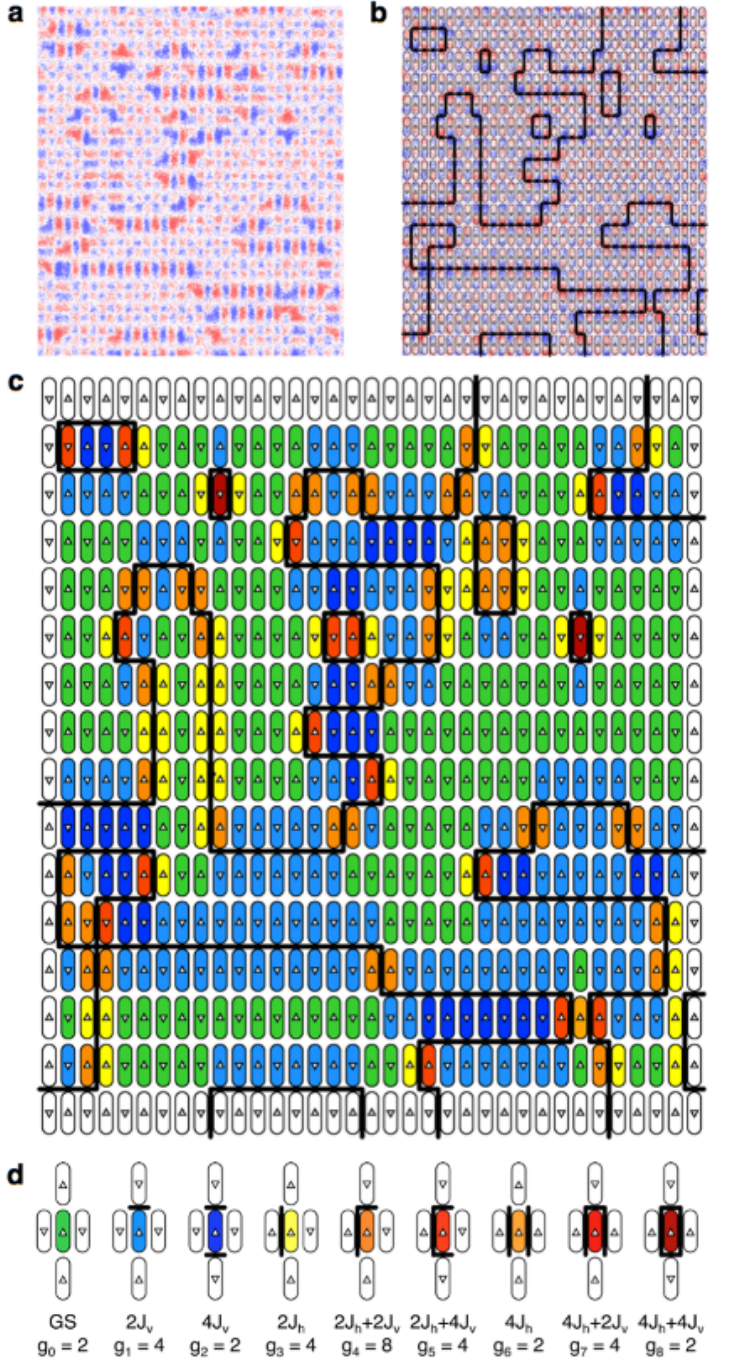}
\end{center}
\caption{Macrospin arrangements imaged by MFM. Analysis of a part of the MFM images recorded showing a portion of the $N=10487$ island array for which the direction of the magnetization was determined. The lateral extent of the entire nano-patterned array was $2\times2$~mm$^2$, corresponding to a total number of islands 40 million. {\bf a} Magnetic force microscopy image showing part of the analyzed array. {\bf b} Schematic highlighting of island contours and the excitation boundaries. {\bf c} The energy state of the islands quantified into the nine different possible energy states  of individual islands with respect to the orientation of their nearest neighbours shown in {\bf d}. The energy states along with their respective degeneracies are listed with respect to the lowest energy ground state assuming that $J_h>J_v$. 
}
\label{fig:MFM}
\end{figure}

%%%%%%%%%%%%%%%%%%%%%%%%%%%%%%%%%%
%\vspace{2ex}\noindent {\bf \large Thermalization and state arrest} \\
%%%%%%%%%%%%%%%%%%%%%%%%%%%%%%%%%%
The thickness regime wherein thermal dynamics of the spins are obtained occurs during the deposition of the magnetic material onto prepatterned substrates as shown by Morgan {\em et al.} \cite{Morgan_NPHYS,Morgan_PRB_2013}. During the deposition, the island thickness (and thereby their volume) becomes gradually larger and the magnetization reversal energy barrier, $E_r$, associated with their shape anisotropy increases. In the initial stages of the island growth this barrier is smaller than the thermal energy enabling the magnetization to spontaneously fluctuate between the two low energy states defined by the shape anisotropy. As the thickness becomes larger the reversal energy increases, eventually overcoming the thermal energy, reaching a threshold where the superparamagnetic behaviour is suppressed as the scale of the combined shape anisotropy reversal energy barrier and the energy landscape of the array are sufficient to lock the magnetization in each of the islands (see Fig. \ref{fig:3dspins}). Subsequently the array can be imaged by magnetic force microscopy (MFM) and the magnetization direction of each island can be determined (see Fig. \ref{fig:MFM}).

During the limited time window where fluctuations occur the fast dynamics allow the array to explore its phase space and achieve an equilibrium condition.
During this dynamic phase before the magnetization becomes frozen the reversal energy barrier and the interaction energies between neighbouring islands is of the order of the thermal energy $k_{\mathrm{B}}T$. For the island sizes and array parameters used for this study the energy values involved correspond to room temperature values for film thicknesses in the $\sim$1 nm regime. Uniform thermal dynamics over the entire array therefore require a well defined, stable thickness of each island in order to minimize randomization effects due to variations in the island thickness and film roughness. Amorphous magnetic materials display soft magnetic properties and a high degree of structural uniformity and are thereby suitable for well defined layers below 1 nm in thickness \cite{ahlberg11}. For this study we therefore choose amorphous Co$_{68}$Fe$_{24}$Zr$_8$ as the island material previously used for creating ultra-thin magnetic layers\cite{Hase_PRB_CoFeZr_XMCD} and well defined nano-patterned multilayered structures \cite{Arnalds:XRMS_2012}. Furthermore, a field imprinted anisotropy can be induced in Co$_{68}$Fe$_{24}$Zr$_8$ enhancing the energy barrier for reversal as the magnetization has settled in a fixed direction \cite{Raanaei:JAP}.

%%%%%%%%%%%%%%%%%%%%%%%%%%%%%%%%%%
%\vspace{2ex}\noindent {\bf \large Thermal ground state ordering} \\
%%%%%%%%%%%%%%%%%%%%%%%%%%%%%%%%%%
Considering magnetostatic interactions in the point dipole approximation between neighbouring islands reveals an  interaction scheme which can be mapped to a ferromagnetic interaction in the vertical direction $J_v>0$ and antiferromagnetic in the horizontal direction $J_h<0$. The lowest energy state of the array is therefore composed of a staggered arrangement of ferromagnetically aligned chains (see Fig. \ref{fig:AFM}). The lowest energy state of an asymmetric two dimensional Ising system is two degenerate, in our case corresponding to a ferromagnetic ordering in the vertical direction and an antiferromagnetic ordering in the horizontal direction. Excitations above the ground state occur through the reversal of a macrospin and can be viewed in the form of boundary walls separating the two possible ground states as illustrated in Fig. \ref{fig:MFM}(c) and the energy state of individual island can be categorized into 9 different energy states of varying degeneracy shown in Fig. \ref{fig:MFM}(d). 
 
\begin{figure}[t]
\begin{center}
  \includegraphics[scale=1]{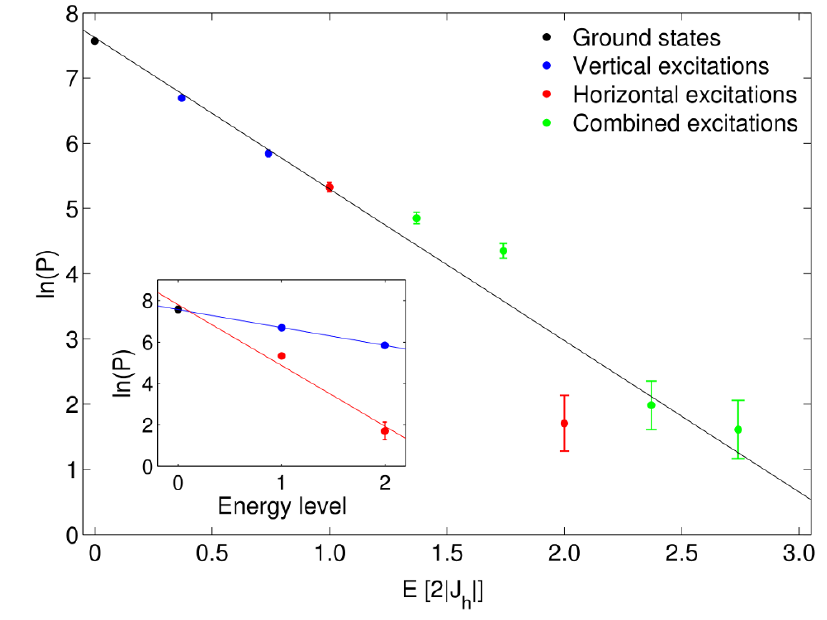}
\end{center}
\caption{Macrospin energy state statistics. The number of observations of each of the 9 energy states illustrated in Fig. \ref{fig:MFM}(d) divided by their degeneracy. The inset shows the number of observed energy states composed exclusively of either vertical excitations (energy levels $2J_v$ and $4J_v$), blue line, or horizontal excitations (energy levels $2J_h$ and $4J_h$), red line. From the ratio of the slopes the ratio of the energy scale between the vertical and  horizontal interaction energies is determined to be $|J_v/J_h|=0.2943$. Considering this ratio the number of observations of the 9 different energy states can be listed as a function of their individual energy in units of $2|J_h|$. The error bars correspond to the square root of the number of observations for each of the energy states.
}
\label{fig:logplot}
\end{figure}

Counting the abundances of excitations composed only of independent vertical or horizontal excitations the relative energy scale between the two directions can be attained. The observed probabilities of these excitations decreases exponentially with increasing energy (inset in Fig. \ref{fig:logplot}) in accordance with a Boltzmann distribution of the states. Determining the ratio of the excitation energies from the inset the energies relating to all energy values of the individual islands can be listed in units of the energy involved with a horizontal excitation $|J_h|$. Within the combined plot (Fig. \ref{fig:logplot}) the observed abundances decrease exponentially establishing the probability for a macrospin to be in an energy state $E$ to be given by a Boltzmann distribution $\sim \exp{(E/k_{\mathrm{B}}T)}$ and that the system can be sufficiently described by a nearest neighbour interaction model.

\begin{figure}[t]
\begin{center}
  \includegraphics[scale=1]{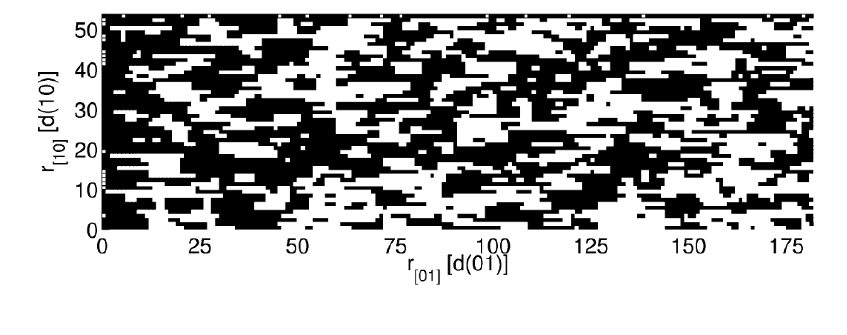}
\end{center}
\caption{Macrospin array domain configuration. The macrospin configuration of the investigated arrays mapped to black and white domains corresponding to the two degenerate ground state of the system. 
}
\label{fig:domain}
\end{figure}

%%%%%%%%%%%%%%%%%%%%%%%%%%%%%%%%%%
%\vspace{2ex}\noindent {\bf \large Order parameter} \\
%%%%%%%%%%%%%%%%%%%%%%%%%%%%%%%%%%
\noindent Identifying the domain walls separating the two degenerate ground states of the array  facilitates the mapping of the magnetic structure of the system as two ordering states, as illustrated in Fig. \ref{fig:3dspins}. The mapping of these two states onto two domain colours for the entire array is shown in Fig. \ref{fig:domain}. Counting the number of islands falling into each of these two domains the order parameter of the array can be obtained. The resulting order parameter for the array, defined by $M=(n_b-n_w)/(n_b+n_w)$ where $n_b$ and $n_w$ correspond to the domain populations of black and white domains, can then be obtained. Truncating the data array to a square shape the array size is reduced to $n = 9828$ out of which $n_b=5283$ islands fall into the {\em black} domain while $n_w = 4545$ islands fall into the {\em white} domain. The resulting value of $M = 0.075\pm0.015$ reveals a slightly higher population of the black domains. Although the statistics of this value are limited by the finite number of observed islands in the array it could be  
indicative of the array being in a state close to, or above $T_c$, since the lack of global order in the macroscopic arrangement can lead to a finite value of the order parameter. The listed uncertainty of $M$ is determined from the square root of the domain populations, $n_w$ and $n_b$. 

\begin{figure}[t]
\begin{center}
  \includegraphics[scale=1]{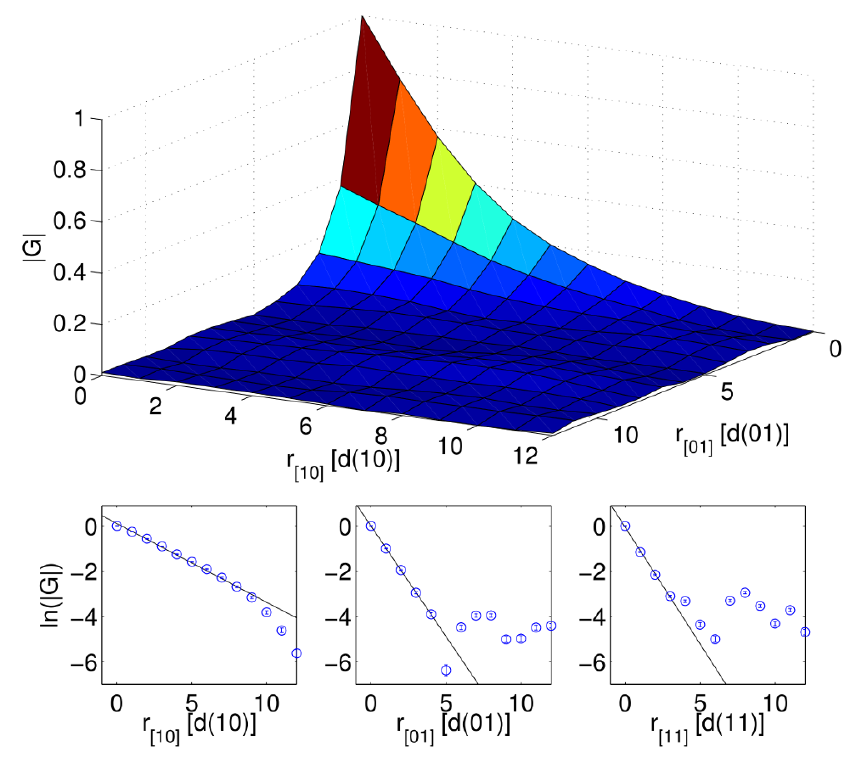}
\end{center}
\caption{Pairwise correlations $G$ for the array. The graph shows the absolute value, $|G|$, for easier mapping of the pairwise correlations for both the ferromagnetic and antiferromagnetic directions. From the slopes the resulting correlation lengths along the different directions is determined to be $\xi_{\mathrm{[10]}} = 2.87$, $\xi_{\mathrm{[01]}} = 1.02$, and $\xi_{\mathrm{[11]}} = 0.97$
}
\label{fig:correlations}
\end{figure}

%%%%%%%%%%%%%%%%%%%%%%%%%%%%%%%%%%
%\vspace{2ex}\noindent {\bf \large Island correlations} \\
%%%%%%%%%%%%%%%%%%%%%%%%%%%%%%%%%%
\noindent By calculating the pairwise correlation between spins, at a distance ${\bf{r}}$, within the experimental array the correlation function for the system, $G({\bf{r}})$, can be determined. 
The resulting correlation array is shown in Fig. \ref{fig:correlations}. 
The preference for an antiferromagnetic arrangement of neighbouring spins in the [10] direction introduces the possibility of negative values in the correlation and alternating positive and negative values along the [10] direction. Figure \ref{fig:correlations} therefore shows the absolute value of the pairwise correlation $|G({\bf{r}})|$ as determined from the array.  
As can be seen in Fig. \ref{fig:correlations} the pairwise correlation follows an exponentially decreasing function as expected for an extended array of Ising like spins at $T>T_c$. 
Furthermore, the asymmetric nature of the array is revealed in the different values of the correlation lengths along the [10] and [01] directions of the array with $\xi_{\mathrm{[10]}} = 2.87$ and $\xi_{\mathrm{[01]}} = 1.02$. The correlation length in the [11] direction, $\xi_{\mathrm{[11]}} = 0.97$, is revealed to be similar to $\xi_{\mathrm{[01]}}$.

\begin{figure}[t]
\begin{center}
  \includegraphics[scale=1]{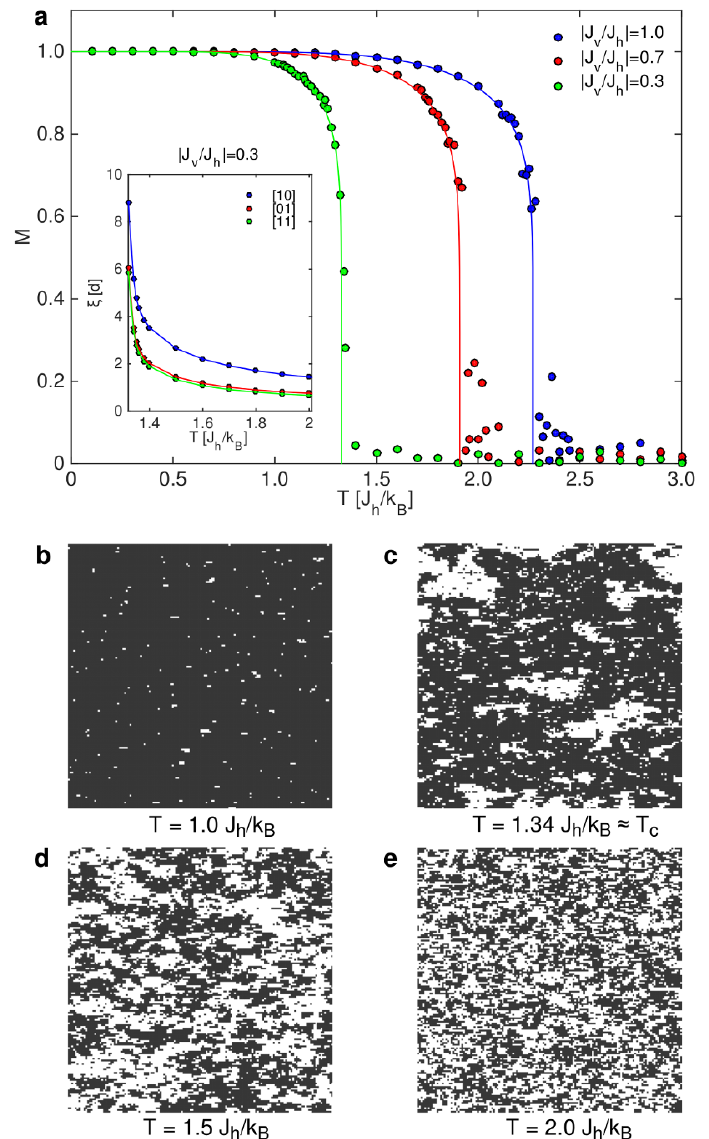} 
\end{center}
\caption{Order parameter versus asymmetry. {\bf a} Results of Monte Carlo simulations showing the order parameter $M$ as a function of temperature, in units of $J_h/k_{\mathrm{B}}$, for different values of $|J_v/J_h|$. Each point in the graph corresponds to an average of several realizations (see methods). As the ratio decreases from the symmetric value of  the $|J_v/J_h| = 1$ the $T_c$ also decreases. The analytical Onsager solution given by equation \eqref{eq:MofT} (solid lines) fits quite well with the obtained numerical data. The inset shows the correlation length $\xi$ as a function of temperature for the ratio $|J_v/J_h| = 0.3$ for the three major crystallographic directions of the array, [10], [01], and [11]. {\bf b} - {\bf e} Representative snapshots of the numerical simulations at temperatures of $1.0J_h/k_{\mathrm{B}}$, $1.34J_h/k_{\mathrm{B}}$, $1.5J_h/k_{\mathrm{B}}$, and  $2.0J_h/k_{\mathrm{B}}$, respectively, for an asymmetry ratio of $|J_v/J_h| = 0.3$. Due to the asymmetry the correlation length in the [10] direction is larger than in the [01] direction stretching out the domains in the [10] direction, similar to what is observed in the experimental data.
}
\label{fig:MC_order_parameter}
\end{figure}

%%%%%%%%%%%%%%%%%%%%%%%%%%%%%%%%%%
%\vspace{2ex}\noindent {\bf \large Ordering in an asymmetric system}\\
%%%%%%%%%%%%%%%%%%%%%%%%%%%%%%%%%%
The role of the ratio of the interaction strengths on the order parameter and specifically the ordering temperature of the array can be investigated using Onsager's solution, as initially done by Chang \cite{Chang:1952} or utilizing numerical calculations such as Monte Carlo simulations, see Fig. \ref{fig:MC_order_parameter}. In the case of an isotropic system, wherein the interaction strength between neighbours, $J$, is the same in the two main lattice directions ([01] and [10]), the ordering temperature is given by $T_c = \frac{2J/k_{\mathrm{B}} }{\ln(1+\sqrt{2})}  \approx 2.269J/k_{\mathrm{B}}$. As $J_v$ decreases with respect to $J_h$ the relative $T_c$ of the system also decreases. Figure \ref{fig:MC_order_parameter} shows results from Monte Carlo simulations for decreasing values of the ratio $|J_v/J_h|$.  The change in $T_c$ corresponds well to what is expected in the two dimensional Ising model as the probability of observing two parallel spins in the [01] direction decreases. 
As can be seen in Fig.~\ref{fig:MC_order_parameter} the results of  numerical simulations correspond well to the analytical solution with small deviations arising from the finite nature of the simulated array. 

Assuming a system in a perfectly ordered ground state at 0~K as the temperature is increased thermal energy is introduced into the system and spins will start to change their directions. The perfect initial order is therefore broken and the system moves towards a thermally disordered state with a reduced order parameter. From numerical simulations this transition can, furthermore, be observed through the pairwise correlation $G({\bf{r}})$ between spins. For the two dimensional  Ising model the correlation function can be written as $G({\bf{r}}) \sim \exp (-\frac{{\bf{r}}}{\xi})$ for $T<T_c$ and $T>T_c$ where $\xi$ is the correlation length affected by temperature and the interaction strength between neighbouring spins. The inset in Fig. \ref{fig:MC_order_parameter} shows the temperature dependence of $\xi$ for the three major directions of the array, [10], [01], and [11] for the asymmetry ratio $|J_v / J_h| = 0.3$ as determined from Monte Carlo calculations. 
As $J_h > J_v$ the corresponding correlation length in the [10] direction is larger than in the other directions while the correlation length in the [01] and [11] directions are similar. Considering the results of the Monte Carlo simulations for an interaction ratio of $|J_v/J_h| = 0.3$ one can see that for temperatures larger than $T_c$ the correlation lengths $\xi_{\mathrm{[10]}}$ and $\xi_{\mathrm{[01]}}$ follow the same trend as the experimental results, i.e. $\xi_{\mathrm{[10]}} > \xi_{\mathrm{[01]}}$ and $\xi_{\mathrm{[01]}}\sim\xi_{\mathrm{[11]}}$. In particular at $T=1.9 J_h/k_{\mathrm{B}}$ the values obtained from the simulations are $\xi_{\mathrm{[10]}} = 1.95$, and [01], $\xi_{\mathrm{[01]}} = 1.06$, $\xi_{\mathrm{[11]}} = 1.08$, similar to the values obtained experimentally. 
These results indicate that in the arrested state the experimental observations can be described with an effective nearest neighbour Ising model. Hence, long range dipolar interactions are not needed to model the present system in the arrested stated. However, dipolar interactions cannot be completely excluded if one desires a full description of the system in any non-arrested state.

%%%%%%%%%%%%%%%%%%%%%%%%%%%%%%%%%%
%\vspace{2ex}\noindent {\bf \large Outlook}\\
%%%%%%%%%%%%%%%%%%%%%%%%%%%%%%%%%%
\noindent Further advances in materials science and magnetic imagining techniques will undoubtedly allow us to follow the thermal evolution of a multitude of different artificial spin systems\cite{Stamps_Heyderman_2013} such as artificial spin ice, one and two dimensional Ising arrays and different frustrated arrangements\cite{Morrison_NJP_2013,Gilbert_Natphys_2014} as they explore their phase space. Such investigations do not merely offer a direct real space probe of the microstate of thermal systems at a local scale but also a direct determination of the dynamics involved. The possibility of determining directly the state of these systems enables the experimental probing of the effect of external variables such as temperature and applied magnetic field. This introduces the possibility of investigating e.g. the two-dimensional Ising model under applied external field directly for which an analytical solution does not exist.

%%%%%%%%%%%%%%%%%%%%%%%%%%%%%%%%%%
\vspace{2ex}\noindent {\bf \large Methods}\\
%%%%%%%%%%%%%%%%%%%%%%%%%%%%%%%%%%
%\noindent{\bf Pattern preparation.} 
The island array was prepared by magnetron sputtering thin film growth onto a pre-patterned fused silica substrate. The 2$\times$2 mm$^2$ pre-patterned array was fabricated by surface conformal nano-imprint lithography \cite{Pierret2010_SCIL,verschuuren_SCIL_2011}. The array was composed of 470 nm $\times$ 160 nm islands arranged in a square lattice with a periodicity of 200 nm along the short axis of the islands ([10] direction) and a periodicity of 500 nm along the [10] direction. The amorphous film was deposited by magnetron sputtering from a compound target at room temperature. Argon was used as the sputtering gas at a pressure of 3.0 mTorr. Before growth the base pressure of the sputtering system was below $2\times10^{-7}$ Pa. Initially a 2 nm thick layer of Al$_{70}$Zr$_{30}$ was grown onto the pre-patterned substrate followed by the growth of a 7 nm thick Co$_{68}$Fe$_{24}$Zr$_{8}$ and finally a 2 nm thick Al$_{70}$Zr$_{30}$ layer to prevent degradation. After patterning the structural quality of the films was investigated by AFM carried out in contact mode using a Nanosurf Mobile S instrument.\\
%\noindent{\bf Magnetic force microscopy imaging.} 
The thermally ordered state of the array was investigated by magnetic force microscopy using a digital instrument dimension 3100. Topography was imaged by height contrast in tapping mode and magnetic micrographs were scanned in lift mode with a lift scan height of 70 nm. In order to reduce the risk of the stray field from the MFM tip altering individual island states the images were recorded using a low magnetic moment Co alloy MFM tip (PPP-LM-MFMR). Repeated scanning of the same area did not alter the state (see supplementary material). \\
%\noindent{\bf Monte-Carlo simulations.} 
The system was modeled as an Ising system, in which every nano-island is considered as a single macrospin with only two possible orientations ($+1$ and $-1$), the simulation cell consists of $150\times 150$ spins. Simulations with both periodic and open boundary conditions were performed. The interactions were only considered between nearest neighbours. The properties of the system were studied via Monte Carlo simulations using the Metropolis-Hastings algorithm. To simulate the process of the spins {\em freezing} at a given configuration and then evolving towards the most stable configuration before any measurement is performed the system is considered to be completely disordered at a temperature higher than its critical temperature. $T_c$. During this thermalization phase no measurements are performed, the system is allowed to evolve in accordance to the usual Metropolis-Hastings algorithm. After a certain number of Monte Carlo steps the system is cooled down and the process is repeated once more. This procedure is then repeated until the desired measurement temperature is reached.

%%%%%%%%%%%%%%%%%%%%%%%%%%%%%%%%%%
\section{Acknowledgements}
%%%%%%%%%%%%%%%%%%%%%%%%%%%%%%%%%%

The authors acknowledge the support of the Knut and Alice Wallenberg Foundation, the Swedish Research Council, and the Swedish Foundation for International Cooperation in Research and Higher Education. UBA acknowledges funding from the Icelandic Research Fund grant nr. 141518-051. 
AB acknowledges the Swedish Research Council (VR) and eSSENCE. The computer simulations were performed on resources provided by the Swedish National Infrastructure for Computing (SNIC) at the National Supercomputer Centre (NSC) and High Performance Computing Center North (HPC2N).

%\bibliography{Two_D_Ising_paper}

\begin{thebibliography}{28}%
\makeatletter
\providecommand \@ifxundefined [1]{%
 \@ifx{#1\undefined}
}%
\providecommand \@ifnum [1]{%
 \ifnum #1\expandafter \@firstoftwo
 \else \expandafter \@secondoftwo
 \fi
}%
\providecommand \@ifx [1]{%
 \ifx #1\expandafter \@firstoftwo
 \else \expandafter \@secondoftwo
 \fi
}%
\providecommand \natexlab [1]{#1}%
\providecommand \enquote  [1]{``#1''}%
\providecommand \bibnamefont  [1]{#1}%
\providecommand \bibfnamefont [1]{#1}%
\providecommand \citenamefont [1]{#1}%
\providecommand \href@noop [0]{\@secondoftwo}%
\providecommand \href [0]{\begingroup \@sanitize@url \@href}%
\providecommand \@href[1]{\@@startlink{#1}\@@href}%
\providecommand \@@href[1]{\endgroup#1\@@endlink}%
\providecommand \@sanitize@url [0]{\catcode `\\12\catcode `\$12\catcode
  `\&12\catcode `\#12\catcode `\^12\catcode `\_12\catcode `\%12\relax}%
\providecommand \@@startlink[1]{}%
\providecommand \@@endlink[0]{}%
\providecommand \url  [0]{\begingroup\@sanitize@url \@url }%
\providecommand \@url [1]{\endgroup\@href {#1}{\urlprefix }}%
\providecommand \urlprefix  [0]{URL }%
\providecommand \Eprint [0]{\href }%
\providecommand \doibase [0]{http://dx.doi.org/}%
\providecommand \selectlanguage [0]{\@gobble}%
\providecommand \bibinfo  [0]{\@secondoftwo}%
\providecommand \bibfield  [0]{\@secondoftwo}%
\providecommand \translation [1]{[#1]}%
\providecommand \BibitemOpen [0]{}%
\providecommand \bibitemStop [0]{}%
\providecommand \bibitemNoStop [0]{.\EOS\space}%
\providecommand \EOS [0]{\spacefactor3000\relax}%
\providecommand \BibitemShut  [1]{\csname bibitem#1\endcsname}%
\let\auto@bib@innerbib\@empty
%</preamble>
\bibitem [{\citenamefont {Ising}(1925)}]{ising}%
  \BibitemOpen
  \bibfield  {author} {\bibinfo {author} {\bibfnamefont {E.}~\bibnamefont
  {Ising}},\ }\bibfield  {title} {\enquote {\bibinfo {title} {Beitrag zur
  {Theorie} des {Ferromagnetismus}},}\ }\href {\doibase 10.1007/BF02980577}
  {\bibfield  {journal} {\bibinfo  {journal} {Z. Phys.}\ }\textbf {\bibinfo
  {volume} {31}},\ \bibinfo {pages} {253} (\bibinfo {year} {1925})}\BibitemShut
  {NoStop}%
\bibitem [{\citenamefont
  {Brush}(1967)}]{History_Ising_model_RevModPhys.39.883}%
  \BibitemOpen
  \bibfield  {author} {\bibinfo {author} {\bibfnamefont {Stephen~G.}\
  \bibnamefont {Brush}},\ }\bibfield  {title} {\enquote {\bibinfo {title}
  {{History of the Lenz-Ising Model}},}\ }\href {\doibase
  10.1103/RevModPhys.39.883} {\bibfield  {journal} {\bibinfo  {journal} {Rev.
  Mod. Phys.}\ }\textbf {\bibinfo {volume} {39}},\ \bibinfo {pages} {883--893}
  (\bibinfo {year} {1967})}\BibitemShut {NoStop}%
\bibitem [{\citenamefont {Loth}\ \emph {et~al.}(2012)\citenamefont {Loth},
  \citenamefont {Baumann}, \citenamefont {Lutz}, \citenamefont {Eigler},\ and\
  \citenamefont {Heinrich}}]{Loth:Science2012}%
  \BibitemOpen
  \bibfield  {author} {\bibinfo {author} {\bibfnamefont {Sebastian}\
  \bibnamefont {Loth}}, \bibinfo {author} {\bibfnamefont {Susanne}\
  \bibnamefont {Baumann}}, \bibinfo {author} {\bibfnamefont {Christopher~P.}\
  \bibnamefont {Lutz}}, \bibinfo {author} {\bibfnamefont {D.~M.}\ \bibnamefont
  {Eigler}}, \ and\ \bibinfo {author} {\bibfnamefont {Andreas~J.}\ \bibnamefont
  {Heinrich}},\ }\bibfield  {title} {\enquote {\bibinfo {title} {Bistability in
  atomic-scale antiferromagnets},}\ }\href {\doibase 10.1126/science.1214131}
  {\bibfield  {journal} {\bibinfo  {journal} {Science}\ }\textbf {\bibinfo
  {volume} {335}},\ \bibinfo {pages} {196--199} (\bibinfo {year} {2012})},\
  \Eprint
  {http://arxiv.org/abs/http://www.sciencemag.org/content/335/6065/196.full.pdf}
  {http://www.sciencemag.org/content/335/6065/196.full.pdf} \BibitemShut
  {NoStop}%
\bibitem [{\citenamefont {Khalil}\ \emph {et~al.}(2012)\citenamefont {Khalil},
  \citenamefont {Sagastegui}, \citenamefont {Li}, \citenamefont {Tahir},
  \citenamefont {Socolar}, \citenamefont {Wiley},\ and\ \citenamefont
  {Yellen}}]{Khalil_NatComm_2012}%
  \BibitemOpen
  \bibfield  {author} {\bibinfo {author} {\bibfnamefont {K.S.}\ \bibnamefont
  {Khalil}}, \bibinfo {author} {\bibfnamefont {A.}~\bibnamefont {Sagastegui}},
  \bibinfo {author} {\bibfnamefont {Y.}~\bibnamefont {Li}}, \bibinfo {author}
  {\bibfnamefont {M.A.}\ \bibnamefont {Tahir}}, \bibinfo {author}
  {\bibfnamefont {J.E.S.}\ \bibnamefont {Socolar}}, \bibinfo {author}
  {\bibfnamefont {B.J.}\ \bibnamefont {Wiley}}, \ and\ \bibinfo {author}
  {\bibfnamefont {B.B.}\ \bibnamefont {Yellen}},\ }\bibfield  {title} {\enquote
  {\bibinfo {title} {{Binary colloidal structures assembled through Ising
  interactions}},}\ }\href
  {http://www.scopus.com/inward/record.url?eid=2-s2.0-84860279506&partnerID=40&md5=256ecfc338ce7fd9ae774e422dc619c2}
  {\bibfield  {journal} {\bibinfo  {journal} {Nature Communications}\ }\textbf
  {\bibinfo {volume} {3}} (\bibinfo {year} {2012})}\BibitemShut {NoStop}%
\bibitem [{\citenamefont {Zhang}\ \emph {et~al.}(2013)\citenamefont {Zhang},
  \citenamefont {Gilbert}, \citenamefont {Nisoli}, \citenamefont {Chern},
  \citenamefont {Erickson}, \citenamefont {O'Brien}, \citenamefont {Leighton},
  \citenamefont {Lammert}, \citenamefont {Crespi},\ and\ \citenamefont
  {Schiffer}}]{Zhang_Nat_2013}%
  \BibitemOpen
  \bibfield  {author} {\bibinfo {author} {\bibfnamefont {S.}~\bibnamefont
  {Zhang}}, \bibinfo {author} {\bibfnamefont {I.}~\bibnamefont {Gilbert}},
  \bibinfo {author} {\bibfnamefont {C.}~\bibnamefont {Nisoli}}, \bibinfo
  {author} {\bibfnamefont {G.-W.}\ \bibnamefont {Chern}}, \bibinfo {author}
  {\bibfnamefont {M.J.}\ \bibnamefont {Erickson}}, \bibinfo {author}
  {\bibfnamefont {L.}~\bibnamefont {O'Brien}}, \bibinfo {author} {\bibfnamefont
  {C.}~\bibnamefont {Leighton}}, \bibinfo {author} {\bibfnamefont {P.E.}\
  \bibnamefont {Lammert}}, \bibinfo {author} {\bibfnamefont {V.H.}\
  \bibnamefont {Crespi}}, \ and\ \bibinfo {author} {\bibfnamefont
  {P.}~\bibnamefont {Schiffer}},\ }\bibfield  {title} {\enquote {\bibinfo
  {title} {Crystallites of magnetic charges in artificial spin ice},}\ }\href
  {http://www.scopus.com/inward/record.url?eid=2-s2.0-84883057884&partnerID=40&md5=5cc3cc71ea64c2560c2b0cb54ac074e1}
  {\bibfield  {journal} {\bibinfo  {journal} {Nature}\ }\textbf {\bibinfo
  {volume} {500}},\ \bibinfo {pages} {553--557} (\bibinfo {year}
  {2013})}\BibitemShut {NoStop}%
\bibitem [{\citenamefont {Peierls}(1936)}]{peierls}%
  \BibitemOpen
  \bibfield  {author} {\bibinfo {author} {\bibfnamefont {R.}~\bibnamefont
  {Peierls}},\ }\bibfield  {title} {\enquote {\bibinfo {title} {{On Ising's
  model of ferromagnetism}},}\ }\href@noop {} {\bibfield  {journal} {\bibinfo
  {journal} {Mathematical Proceedings of the Cambridge Philosophical Society}\
  }\textbf {\bibinfo {volume} {32}},\ \bibinfo {pages} {477--481} (\bibinfo
  {year} {1936})}\BibitemShut {NoStop}%
\bibitem [{\citenamefont {Onsager}(1944)}]{Onsager_2d_Ising}%
  \BibitemOpen
  \bibfield  {author} {\bibinfo {author} {\bibfnamefont {Lars}\ \bibnamefont
  {Onsager}},\ }\bibfield  {title} {\enquote {\bibinfo {title} {{Crystal
  Statistics. 1. A Two-Dimensional Model with an Order-Disorder Transition}},}\
  }\href@noop {} {\bibfield  {journal} {\bibinfo  {journal} {Physical Review}\
  }\textbf {\bibinfo {volume} {65}},\ \bibinfo {pages} {117} (\bibinfo {year}
  {1944})}\BibitemShut {NoStop}%
\bibitem [{\citenamefont {Yang}(1952)}]{Yang:1952}%
  \BibitemOpen
  \bibfield  {author} {\bibinfo {author} {\bibfnamefont {C.~N.}\ \bibnamefont
  {Yang}},\ }\bibfield  {title} {\enquote {\bibinfo {title} {{The Spontaneous
  Magnetization of a Two-Dimensional Ising Model}},}\ }\href {\doibase
  http://dx.doi.org/10.1103/PhysRev.85.808} {\bibfield  {journal} {\bibinfo
  {journal} {Phys. Rev.}\ }\textbf {\bibinfo {volume} {85}},\ \bibinfo {pages}
  {808} (\bibinfo {year} {1952})}\BibitemShut {NoStop}%
\bibitem [{\citenamefont {Imre}\ \emph {et~al.}(2006)\citenamefont {Imre},
  \citenamefont {Csaba}, \citenamefont {Ji}, \citenamefont {Orlov},
  \citenamefont {Bernstein},\ and\ \citenamefont {Porod}}]{Imre_Science_2006}%
  \BibitemOpen
  \bibfield  {author} {\bibinfo {author} {\bibfnamefont {A.}~\bibnamefont
  {Imre}}, \bibinfo {author} {\bibfnamefont {G.}~\bibnamefont {Csaba}},
  \bibinfo {author} {\bibfnamefont {L.}~\bibnamefont {Ji}}, \bibinfo {author}
  {\bibfnamefont {A.}~\bibnamefont {Orlov}}, \bibinfo {author} {\bibfnamefont
  {G.~H.}\ \bibnamefont {Bernstein}}, \ and\ \bibinfo {author} {\bibfnamefont
  {W.}~\bibnamefont {Porod}},\ }\bibfield  {title} {\enquote {\bibinfo {title}
  {Majority logic gate for magnetic quantum-dot cellular automata},}\ }\href
  {\doibase 10.1126/science.1120506} {\bibfield  {journal} {\bibinfo  {journal}
  {Science}\ }\textbf {\bibinfo {volume} {311}},\ \bibinfo {pages} {205--208}
  (\bibinfo {year} {2006})}\BibitemShut {NoStop}%
\bibitem [{\citenamefont {Cowburn}\ and\ \citenamefont
  {Welland}(2000)}]{Cowburn_Science_2000}%
  \BibitemOpen
  \bibfield  {author} {\bibinfo {author} {\bibfnamefont {R.~P.}\ \bibnamefont
  {Cowburn}}\ and\ \bibinfo {author} {\bibfnamefont {M.~E.}\ \bibnamefont
  {Welland}},\ }\bibfield  {title} {\enquote {\bibinfo {title} {Room
  temperature magnetic quantum cellular automata},}\ }\href@noop {} {\bibfield
  {journal} {\bibinfo  {journal} {Science}\ }\textbf {\bibinfo {volume}
  {287}},\ \bibinfo {pages} {1466} (\bibinfo {year} {2000})}\BibitemShut
  {NoStop}%
\bibitem [{\citenamefont {Wang}\ \emph {et~al.}(2006)\citenamefont {Wang},
  \citenamefont {Nisoli}, \citenamefont {Freitas}, \citenamefont {Li},
  \citenamefont {McConville}, \citenamefont {Cooley}, \citenamefont {Lund},
  \citenamefont {Samarth}, \citenamefont {Leighton}, \citenamefont {Crespi},\
  and\ \citenamefont {Schiffer}}]{wang_nature}%
  \BibitemOpen
  \bibfield  {author} {\bibinfo {author} {\bibfnamefont {R.~F.}\ \bibnamefont
  {Wang}}, \bibinfo {author} {\bibfnamefont {C.}~\bibnamefont {Nisoli}},
  \bibinfo {author} {\bibfnamefont {R.~S.}\ \bibnamefont {Freitas}}, \bibinfo
  {author} {\bibfnamefont {J.}~\bibnamefont {Li}}, \bibinfo {author}
  {\bibfnamefont {W.}~\bibnamefont {McConville}}, \bibinfo {author}
  {\bibfnamefont {B.~J.}\ \bibnamefont {Cooley}}, \bibinfo {author}
  {\bibfnamefont {M.~S.}\ \bibnamefont {Lund}}, \bibinfo {author}
  {\bibfnamefont {N.}~\bibnamefont {Samarth}}, \bibinfo {author} {\bibfnamefont
  {C.}~\bibnamefont {Leighton}}, \bibinfo {author} {\bibfnamefont {V.~H.}\
  \bibnamefont {Crespi}}, \ and\ \bibinfo {author} {\bibfnamefont
  {P.}~\bibnamefont {Schiffer}},\ }\bibfield  {title} {\enquote {\bibinfo
  {title} {{Artificial `spin ice' in a geometrically frustrated lattice of
  nanoscale ferromagnetic islands}},}\ }\href {\doibase 10.1038/nature04447}
  {\bibfield  {journal} {\bibinfo  {journal} {Nature}\ }\textbf {\bibinfo
  {volume} {439}},\ \bibinfo {pages} {303--306} (\bibinfo {year}
  {2006})}\BibitemShut {NoStop}%
\bibitem [{\citenamefont {Nisoli}\ \emph {et~al.}(2013)\citenamefont {Nisoli},
  \citenamefont {Moessner},\ and\ \citenamefont {Schiffer}}]{Nisoli_Review}%
  \BibitemOpen
  \bibfield  {author} {\bibinfo {author} {\bibfnamefont {Cristiano}\
  \bibnamefont {Nisoli}}, \bibinfo {author} {\bibfnamefont {Roderich}\
  \bibnamefont {Moessner}}, \ and\ \bibinfo {author} {\bibfnamefont {Peter}\
  \bibnamefont {Schiffer}},\ }\bibfield  {title} {\enquote {\bibinfo {title}
  {\textit{Colloquium} : Artificial spin ice: Designing and imaging magnetic
  frustration},}\ }\href {\doibase 10.1103/RevModPhys.85.1473} {\bibfield
  {journal} {\bibinfo  {journal} {Rev. Mod. Phys.}\ }\textbf {\bibinfo {volume}
  {85}},\ \bibinfo {pages} {1473--1490} (\bibinfo {year} {2013})}\BibitemShut
  {NoStop}%
\bibitem [{\citenamefont {Morgan}\ \emph {et~al.}(2011)\citenamefont {Morgan},
  \citenamefont {Stein}, \citenamefont {Langridge},\ and\ \citenamefont
  {Marrows}}]{Morgan_NPHYS}%
  \BibitemOpen
  \bibfield  {author} {\bibinfo {author} {\bibfnamefont {J.~P.}\ \bibnamefont
  {Morgan}}, \bibinfo {author} {\bibfnamefont {A.}~\bibnamefont {Stein}},
  \bibinfo {author} {\bibfnamefont {S.}~\bibnamefont {Langridge}}, \ and\
  \bibinfo {author} {\bibfnamefont {C.~H.}\ \bibnamefont {Marrows}},\
  }\bibfield  {title} {\enquote {\bibinfo {title} {Thermal ground-state
  ordering and elementary excitations in artificial magnetic square ice},}\
  }\href {\doibase doi:10.1038/nphys1853} {\bibfield  {journal} {\bibinfo
  {journal} {Nat. Phys.}\ }\textbf {\bibinfo {volume} {7}},\ \bibinfo {pages}
  {75} (\bibinfo {year} {2011})}\BibitemShut {NoStop}%
\bibitem [{\citenamefont {Mengotti}\ \emph {et~al.}(2011)\citenamefont
  {Mengotti}, \citenamefont {Heyderman}, \citenamefont {Fraile~Rodr\'{i}guez},
  \citenamefont {Nolting}, \citenamefont {H\"{u}gli},\ and\ \citenamefont
  {Braun}}]{mengotti:NPHYS2011}%
  \BibitemOpen
  \bibfield  {author} {\bibinfo {author} {\bibfnamefont {E.}~\bibnamefont
  {Mengotti}}, \bibinfo {author} {\bibfnamefont {L.~J.}\ \bibnamefont
  {Heyderman}}, \bibinfo {author} {\bibfnamefont {A}~\bibnamefont
  {Fraile~Rodr\'{i}guez}}, \bibinfo {author} {\bibfnamefont {F.}~\bibnamefont
  {Nolting}}, \bibinfo {author} {\bibfnamefont {R.~V.}\ \bibnamefont
  {H\"{u}gli}}, \ and\ \bibinfo {author} {\bibfnamefont {H.~B.}\ \bibnamefont
  {Braun}},\ }\bibfield  {title} {\enquote {\bibinfo {title} {Real-space
  observation of emergent magnetic monopoles and associated dirac strings in
  artificial kagome spin ice},}\ }\href {\doibase 10.1038/nphys1794} {\bibfield
   {journal} {\bibinfo  {journal} {Nat. Phys.}\ }\textbf {\bibinfo {volume}
  {7}},\ \bibinfo {pages} {68} (\bibinfo {year} {2011})}\BibitemShut {NoStop}%
\bibitem [{\citenamefont {Arnalds}\ \emph
  {et~al.}(2012{\natexlab{a}})\citenamefont {Arnalds}, \citenamefont {Farhan},
  \citenamefont {Chopdekar}, \citenamefont {Kapaklis}, \citenamefont {Balan},
  \citenamefont {Papaioannou}, \citenamefont {Ahlberg}, \citenamefont
  {Nolting}, \citenamefont {Heyderman},\ and\ \citenamefont
  {Hj\"{o}rvarsson}}]{KagomeAPL}%
  \BibitemOpen
  \bibfield  {author} {\bibinfo {author} {\bibfnamefont {U.~B.}\ \bibnamefont
  {Arnalds}}, \bibinfo {author} {\bibfnamefont {A.}~\bibnamefont {Farhan}},
  \bibinfo {author} {\bibfnamefont {R.~V.}\ \bibnamefont {Chopdekar}}, \bibinfo
  {author} {\bibfnamefont {V.}~\bibnamefont {Kapaklis}}, \bibinfo {author}
  {\bibfnamefont {A.}~\bibnamefont {Balan}}, \bibinfo {author} {\bibfnamefont
  {E.~Th.}\ \bibnamefont {Papaioannou}}, \bibinfo {author} {\bibfnamefont
  {M.}~\bibnamefont {Ahlberg}}, \bibinfo {author} {\bibfnamefont
  {F.}~\bibnamefont {Nolting}}, \bibinfo {author} {\bibfnamefont {L.~J.}\
  \bibnamefont {Heyderman}}, \ and\ \bibinfo {author} {\bibfnamefont
  {B.}~\bibnamefont {Hj\"{o}rvarsson}},\ }\bibfield  {title} {\enquote
  {\bibinfo {title} {Thermalized ground state of artificial kagome spin ice
  building blocks},}\ }\href {\doibase 10.1063/1.4751844} {\bibfield  {journal}
  {\bibinfo  {journal} {Appl. Phys. Lett.}\ }\textbf {\bibinfo {volume}
  {101}},\ \bibinfo {pages} {112404} (\bibinfo {year}
  {2012}{\natexlab{a}})}\BibitemShut {NoStop}%
\bibitem [{\citenamefont {Kapaklis}\ \emph {et~al.}(2014)\citenamefont
  {Kapaklis}, \citenamefont {Arnalds}, \citenamefont {Farhan}, \citenamefont
  {Chopdekar}, \citenamefont {Balan}, \citenamefont {A}, \citenamefont
  {Scholl}, \citenamefont {Heyderman},\ and\ \citenamefont
  {Hj\"{o}rvarsson}}]{Kapaklis:NatNano:2014}%
  \BibitemOpen
  \bibfield  {author} {\bibinfo {author} {\bibfnamefont {V.}~\bibnamefont
  {Kapaklis}}, \bibinfo {author} {\bibfnamefont {U.~B.}\ \bibnamefont
  {Arnalds}}, \bibinfo {author} {\bibfnamefont {A.}~\bibnamefont {Farhan}},
  \bibinfo {author} {\bibfnamefont {R.~V.}\ \bibnamefont {Chopdekar}}, \bibinfo
  {author} {\bibfnamefont {A.}~\bibnamefont {Balan}}, \bibinfo {author}
  {\bibnamefont {A}}, \bibinfo {author} {\bibnamefont {Scholl}}, \bibinfo
  {author} {\bibfnamefont {L.~J.}\ \bibnamefont {Heyderman}}, \ and\ \bibinfo
  {author} {\bibfnamefont {B.}~\bibnamefont {Hj\"{o}rvarsson}},\ }\bibfield
  {title} {\enquote {\bibinfo {title} {Thermal fluctuations in artificial spin
  ice},}\ }\href {\doibase doi:10.1038/nnano.2014.104} {\bibfield  {journal}
  {\bibinfo  {journal} {Nature Nanotechnology}\ }\textbf {\bibinfo {volume}
  {9}},\ \bibinfo {pages} {514} (\bibinfo {year} {2014})}\BibitemShut {NoStop}%
\bibitem [{\citenamefont {Daunheimer}\ \emph {et~al.}(2011)\citenamefont
  {Daunheimer}, \citenamefont {Petrova}, \citenamefont {Tchernyshyov},\ and\
  \citenamefont {Cumings}}]{Daunheimer:PRL:2011}%
  \BibitemOpen
  \bibfield  {author} {\bibinfo {author} {\bibfnamefont {S.~A.}\ \bibnamefont
  {Daunheimer}}, \bibinfo {author} {\bibfnamefont {O.}~\bibnamefont {Petrova}},
  \bibinfo {author} {\bibfnamefont {O.}~\bibnamefont {Tchernyshyov}}, \ and\
  \bibinfo {author} {\bibfnamefont {J.}~\bibnamefont {Cumings}},\ }\bibfield
  {title} {\enquote {\bibinfo {title} {Reducing disorder in artificial kagome
  ice},}\ }\href {\doibase 10.1103/PhysRevLett.107.167201} {\bibfield
  {journal} {\bibinfo  {journal} {Phys. Rev. Lett.}\ }\textbf {\bibinfo
  {volume} {107}} (\bibinfo {year} {2011}),\
  10.1103/PhysRevLett.107.167201}\BibitemShut {NoStop}%
\bibitem [{\citenamefont {Morgan}\ \emph {et~al.}(2013)\citenamefont {Morgan},
  \citenamefont {Akerman}, \citenamefont {Stein}, \citenamefont {Phatak},
  \citenamefont {Evans}, \citenamefont {Langridge},\ and\ \citenamefont
  {Marrows}}]{Morgan_PRB_2013}%
  \BibitemOpen
  \bibfield  {author} {\bibinfo {author} {\bibfnamefont {Jason~P.}\
  \bibnamefont {Morgan}}, \bibinfo {author} {\bibfnamefont {Johanna}\
  \bibnamefont {Akerman}}, \bibinfo {author} {\bibfnamefont {Aaron}\
  \bibnamefont {Stein}}, \bibinfo {author} {\bibfnamefont {Charudatta}\
  \bibnamefont {Phatak}}, \bibinfo {author} {\bibfnamefont {R.~M.~L.}\
  \bibnamefont {Evans}}, \bibinfo {author} {\bibfnamefont {Sean}\ \bibnamefont
  {Langridge}}, \ and\ \bibinfo {author} {\bibfnamefont {Christopher~H.}\
  \bibnamefont {Marrows}},\ }\bibfield  {title} {\enquote {\bibinfo {title}
  {Real and effective thermal equilibrium in artificial square spin ices},}\
  }\href {\doibase 10.1103/PhysRevB.87.024405} {\bibfield  {journal} {\bibinfo
  {journal} {Phys. Rev. B}\ }\textbf {\bibinfo {volume} {87}},\ \bibinfo
  {pages} {024405} (\bibinfo {year} {2013})}\BibitemShut {NoStop}%
\bibitem [{\citenamefont {Ahlberg}\ \emph {et~al.}(2011)\citenamefont
  {Ahlberg}, \citenamefont {Andersson},\ and\ \citenamefont
  {Hj\"orvarsson}}]{ahlberg11}%
  \BibitemOpen
  \bibfield  {author} {\bibinfo {author} {\bibfnamefont {M.}~\bibnamefont
  {Ahlberg}}, \bibinfo {author} {\bibfnamefont {G.}~\bibnamefont {Andersson}},
  \ and\ \bibinfo {author} {\bibfnamefont {B.}~\bibnamefont {Hj\"orvarsson}},\
  }\bibfield  {title} {\enquote {\bibinfo {title} {Two-dimensional
  $\mathit{XY}$-like amorphous
  {Co}${}_{68}${Fe}${}_{24}${Zr}${}_{8}$/{Al}${}_{70}${Zr}${}_{30}$
  multilayers},}\ }\href {\doibase 10.1103/PhysRevB.83.224404} {\bibfield
  {journal} {\bibinfo  {journal} {Phys. Rev. B}\ }\textbf {\bibinfo {volume}
  {83}},\ \bibinfo {pages} {224404} (\bibinfo {year} {2011})}\BibitemShut
  {NoStop}%
\bibitem [{\citenamefont {Hase}\ \emph {et~al.}(2009)\citenamefont {Hase},
  \citenamefont {Raanaei}, \citenamefont {Lidbaum}, \citenamefont
  {S\'anchez-Hanke}, \citenamefont {Wilkins}, \citenamefont {Leifer},\ and\
  \citenamefont {Hj\"orvarsson}}]{Hase_PRB_CoFeZr_XMCD}%
  \BibitemOpen
  \bibfield  {author} {\bibinfo {author} {\bibfnamefont {Thomas}\ \bibnamefont
  {Hase}}, \bibinfo {author} {\bibfnamefont {Hossein}\ \bibnamefont {Raanaei}},
  \bibinfo {author} {\bibfnamefont {Hans}\ \bibnamefont {Lidbaum}}, \bibinfo
  {author} {\bibfnamefont {Cecilia}\ \bibnamefont {S\'anchez-Hanke}}, \bibinfo
  {author} {\bibfnamefont {Stuart}\ \bibnamefont {Wilkins}}, \bibinfo {author}
  {\bibfnamefont {Klaus}\ \bibnamefont {Leifer}}, \ and\ \bibinfo {author}
  {\bibfnamefont {Bj\"orgvin}\ \bibnamefont {Hj\"orvarsson}},\ }\bibfield
  {title} {\enquote {\bibinfo {title} {Spin and orbital moment in amorphous
  {Co}$_{68}${Fe}$_{24}${Zr}$_{8}$ layers},}\ }\href {\doibase
  10.1103/PhysRevB.80.134402} {\bibfield  {journal} {\bibinfo  {journal} {Phys.
  Rev. B}\ }\textbf {\bibinfo {volume} {80}},\ \bibinfo {pages} {134402}
  (\bibinfo {year} {2009})}\BibitemShut {NoStop}%
\bibitem [{\citenamefont {Arnalds}\ \emph
  {et~al.}(2012{\natexlab{b}})\citenamefont {Arnalds}, \citenamefont {Hase},
  \citenamefont {Papaioannou}, \citenamefont {Raanaei}, \citenamefont
  {Abrudan}, \citenamefont {Charlton}, \citenamefont {Langridge},\ and\
  \citenamefont {Hj\"orvarsson}}]{Arnalds:XRMS_2012}%
  \BibitemOpen
  \bibfield  {author} {\bibinfo {author} {\bibfnamefont {Unnar~B.}\
  \bibnamefont {Arnalds}}, \bibinfo {author} {\bibfnamefont {Thomas P.~A.}\
  \bibnamefont {Hase}}, \bibinfo {author} {\bibfnamefont {Evangelos~Th.}\
  \bibnamefont {Papaioannou}}, \bibinfo {author} {\bibfnamefont {Hossein}\
  \bibnamefont {Raanaei}}, \bibinfo {author} {\bibfnamefont {Radu}\
  \bibnamefont {Abrudan}}, \bibinfo {author} {\bibfnamefont {Timothy~R.}\
  \bibnamefont {Charlton}}, \bibinfo {author} {\bibfnamefont {Sean}\
  \bibnamefont {Langridge}}, \ and\ \bibinfo {author} {\bibfnamefont
  {Bj\"orgvin}\ \bibnamefont {Hj\"orvarsson}},\ }\bibfield  {title} {\enquote
  {\bibinfo {title} {X-ray resonant magnetic scattering from patterned
  multilayers},}\ }\href {\doibase 10.1103/PhysRevB.86.064426} {\bibfield
  {journal} {\bibinfo  {journal} {Phys. Rev. B}\ }\textbf {\bibinfo {volume}
  {86}},\ \bibinfo {pages} {064426} (\bibinfo {year}
  {2012}{\natexlab{b}})}\BibitemShut {NoStop}%
\bibitem [{\citenamefont {Raanaei}\ \emph {et~al.}(2009)\citenamefont
  {Raanaei}, \citenamefont {Nguyen}, \citenamefont {Andersson}, \citenamefont
  {Lidbaum}, \citenamefont {Korelis}, \citenamefont {Leifer},\ and\
  \citenamefont {Hj\"{o}rvarsson}}]{Raanaei:JAP}%
  \BibitemOpen
  \bibfield  {author} {\bibinfo {author} {\bibfnamefont {H.}~\bibnamefont
  {Raanaei}}, \bibinfo {author} {\bibfnamefont {H.}~\bibnamefont {Nguyen}},
  \bibinfo {author} {\bibfnamefont {G.}~\bibnamefont {Andersson}}, \bibinfo
  {author} {\bibfnamefont {H.}~\bibnamefont {Lidbaum}}, \bibinfo {author}
  {\bibfnamefont {P.}~\bibnamefont {Korelis}}, \bibinfo {author} {\bibfnamefont
  {K.}~\bibnamefont {Leifer}}, \ and\ \bibinfo {author} {\bibfnamefont
  {B.}~\bibnamefont {Hj\"{o}rvarsson}},\ }\bibfield  {title} {\enquote
  {\bibinfo {title} {Imprinting layer specific magnetic anisotropies in
  amorphous multilayers},}\ }\href {\doibase
  http://dx.doi.org/10.1063/1.3169523} {\bibfield  {journal} {\bibinfo
  {journal} {Journal of Applied Physics}\ }\textbf {\bibinfo {volume} {106}},\
  \bibinfo {eid} {023918} (\bibinfo {year} {2009})}\BibitemShut {NoStop}%
\bibitem [{\citenamefont {Chang}(1952)}]{Chang:1952}%
  \BibitemOpen
  \bibfield  {author} {\bibinfo {author} {\bibfnamefont {C.~H.}\ \bibnamefont
  {Chang}},\ }\bibfield  {title} {\enquote {\bibinfo {title} {{The Spontaneous
  Magnetization of a Two-Dimensional Rectangular Ising Model}},}\ }\href
  {\doibase http://dx.doi.org/10.1103/PhysRev.88.1422} {\bibfield  {journal}
  {\bibinfo  {journal} {Phys. Rev.}\ }\textbf {\bibinfo {volume} {88}},\
  \bibinfo {pages} {1422} (\bibinfo {year} {1952})}\BibitemShut {NoStop}%
\bibitem [{\citenamefont {Heyderman}\ and\ \citenamefont
  {Stamps}(2013)}]{Stamps_Heyderman_2013}%
  \BibitemOpen
  \bibfield  {author} {\bibinfo {author} {\bibfnamefont {L.J.}\ \bibnamefont
  {Heyderman}}\ and\ \bibinfo {author} {\bibfnamefont {R.L.}\ \bibnamefont
  {Stamps}},\ }\bibfield  {title} {\enquote {\bibinfo {title} {Artificial
  ferroic systems: Novel functionality from structure, interactions and
  dynamics},}\ }\href
  {http://www.scopus.com/inward/record.url?eid=2-s2.0-84882734272&partnerID=40&md5=77223903ed6053fdcb9b4157c440fa85}
  {\bibfield  {journal} {\bibinfo  {journal} {Journal of Physics Condensed
  Matter}\ }\textbf {\bibinfo {volume} {25}} (\bibinfo {year}
  {2013})}\BibitemShut {NoStop}%
\bibitem [{\citenamefont {Morrison}\ \emph {et~al.}(2013)\citenamefont
  {Morrison}, \citenamefont {Nelson},\ and\ \citenamefont
  {Nisoli}}]{Morrison_NJP_2013}%
  \BibitemOpen
  \bibfield  {author} {\bibinfo {author} {\bibfnamefont {M.J.}\ \bibnamefont
  {Morrison}}, \bibinfo {author} {\bibfnamefont {T.R.}\ \bibnamefont {Nelson}},
  \ and\ \bibinfo {author} {\bibfnamefont {C.}~\bibnamefont {Nisoli}},\
  }\bibfield  {title} {\enquote {\bibinfo {title} {Unhappy vertices in
  artificial spin ice: New degeneracies from vertex frustration},}\ }\href
  {http://www.scopus.com/inward/record.url?eid=2-s2.0-84876748564&partnerID=40&md5=36c756ef69024fd6e0d7a7bc4eee8aa4}
  {\bibfield  {journal} {\bibinfo  {journal} {New Journal of Physics}\ }\textbf
  {\bibinfo {volume} {15}} (\bibinfo {year} {2013})},\ \bibinfo {note} {cited
  By (since 1996)1}\BibitemShut {NoStop}%
\bibitem [{\citenamefont {Gilbert}\ \emph {et~al.}(2014)\citenamefont
  {Gilbert}, \citenamefont {Chern}, \citenamefont {Zhang}, \citenamefont
  {O'Brien}, \citenamefont {Fore}, \citenamefont {Nisoli},\ and\ \citenamefont
  {Schiffer}}]{Gilbert_Natphys_2014}%
  \BibitemOpen
  \bibfield  {author} {\bibinfo {author} {\bibfnamefont {Ian}\ \bibnamefont
  {Gilbert}}, \bibinfo {author} {\bibfnamefont {Gia-Wei}\ \bibnamefont
  {Chern}}, \bibinfo {author} {\bibfnamefont {Sheng}\ \bibnamefont {Zhang}},
  \bibinfo {author} {\bibfnamefont {Liam}\ \bibnamefont {O'Brien}}, \bibinfo
  {author} {\bibfnamefont {Bryce}\ \bibnamefont {Fore}}, \bibinfo {author}
  {\bibfnamefont {Cristiano}\ \bibnamefont {Nisoli}}, \ and\ \bibinfo {author}
  {\bibfnamefont {Peter}\ \bibnamefont {Schiffer}},\ }\bibfield  {title}
  {\enquote {\bibinfo {title} {Emergent ice rule and magnetic charge screening
  from vertex frustration in artificial spin ice},}\ }\href {\doibase
  10.1038/NPHYS3037} {\bibfield  {journal} {\bibinfo  {journal} {Nat. Phys.}\
  }\textbf {\bibinfo {volume} {10}},\ \bibinfo {pages} {670} (\bibinfo {year}
  {2014})}\BibitemShut {NoStop}%
\bibitem [{\citenamefont {Pierret}\ \emph {et~al.}(2010)\citenamefont
  {Pierret}, \citenamefont {Hocevar}, \citenamefont {Diedenhofen},
  \citenamefont {Algra}, \citenamefont {Vlieg}, \citenamefont {Timmering},
  \citenamefont {Verschuuren}, \citenamefont {Immink}, \citenamefont
  {Verheijen},\ and\ \citenamefont {Bakkers}}]{Pierret2010_SCIL}%
  \BibitemOpen
  \bibfield  {author} {\bibinfo {author} {\bibfnamefont {A.}~\bibnamefont
  {Pierret}}, \bibinfo {author} {\bibfnamefont {M.}~\bibnamefont {Hocevar}},
  \bibinfo {author} {\bibfnamefont {S.L.}\ \bibnamefont {Diedenhofen}},
  \bibinfo {author} {\bibfnamefont {R.E.}\ \bibnamefont {Algra}}, \bibinfo
  {author} {\bibfnamefont {E.}~\bibnamefont {Vlieg}}, \bibinfo {author}
  {\bibfnamefont {E.C.}\ \bibnamefont {Timmering}}, \bibinfo {author}
  {\bibfnamefont {M.A.}\ \bibnamefont {Verschuuren}}, \bibinfo {author}
  {\bibfnamefont {G.W.G.}\ \bibnamefont {Immink}}, \bibinfo {author}
  {\bibfnamefont {M.A.}\ \bibnamefont {Verheijen}}, \ and\ \bibinfo {author}
  {\bibfnamefont {E.P.A.M.}\ \bibnamefont {Bakkers}},\ }\bibfield  {title}
  {\enquote {\bibinfo {title} {Generic nano-imprint process for fabrication of
  nanowire arrays},}\ }\href
  {http://www.scopus.com/inward/record.url?eid=2-s2.0-75249091313&partnerID=40&md5=85fa6d96d17198ae3bf67f61203b91c1}
  {\bibfield  {journal} {\bibinfo  {journal} {Nanotechnology}\ }\textbf
  {\bibinfo {volume} {21}} (\bibinfo {year} {2010})}\BibitemShut {NoStop}%
\bibitem [{\citenamefont {Verschuuren}\ \emph {et~al.}(2011)\citenamefont
  {Verschuuren}, \citenamefont {Gerlach}, \citenamefont {van Sprang},\ and\
  \citenamefont {Polman}}]{verschuuren_SCIL_2011}%
  \BibitemOpen
  \bibfield  {author} {\bibinfo {author} {\bibfnamefont {M.~A.}\ \bibnamefont
  {Verschuuren}}, \bibinfo {author} {\bibfnamefont {P.}~\bibnamefont
  {Gerlach}}, \bibinfo {author} {\bibfnamefont {H.~A.}\ \bibnamefont {van
  Sprang}}, \ and\ \bibinfo {author} {\bibfnamefont {A}~\bibnamefont
  {Polman}},\ }\bibfield  {title} {\enquote {\bibinfo {title} {Improved
  performance of polarization-stable {VCSELs} by monolithic sub-wavelength
  gratings produced by soft nano-imprint lithography},}\ }\href {\doibase
  doi:10.1088/0957-4484/22/50/505201} {\bibfield  {journal} {\bibinfo
  {journal} {Nanotechnology}\ }\textbf {\bibinfo {volume} {22}},\ \bibinfo
  {pages} {505201} (\bibinfo {year} {2011})}\BibitemShut {NoStop}%
\end{thebibliography}

\end{document}